\documentclass{svproc}
\usepackage{url}
\usepackage{amsmath}
\usepackage{amssymb}
\usepackage[space]{cite}

\usepackage{xcolor}

\newcommand{\Tr}{\text{Tr}}

\begin{document}
\mainmatter              % start of a contribution
\title{Entanglement of free-fermion systems, signal processing and algebraic combinatorics}
\titlerunning{Entanglement of free fermions}  % abbreviated title (for running head)
\author{Pierre-Antoine Bernard\inst{1} \and Nicolas Cramp\'e\inst{2} \and
Rafael I. Nepomechie\inst{3} \and \\[0.1in] Gilles Parez \inst{1} \and Luc Vinet \inst{1,4}}
\authorrunning{Pierre-Antoine Bernard et al.} % abbreviated author list (for running head)
\tocauthor{Pierre-Antoine Bernard, Nicolas Cramp\'e, Rafael Nepomechie, Gilles Parez, Luc Vinet}
\institute{Centre de Recherches Math\'ematiques and D\'epartement de Physique,\\ Universit\'e de Montr\'eal, P.O. Box 6128, \\
\small Centre-ville Station, Montr\'eal (Qu\'ebec), H3C 3J7, Canada.\\
\email{pierre-antoine.bernard@umontreal.ca, gilles.parez@umontreal.ca, luc.vinet@umontreal.ca}
\and
Institut Denis-Poisson CNRS/UMR 7013\\ - Universit\'e de Tours - Universit\'e
d'Orl\'eans,\\
\small~Parc de Grandmont, 37200 Tours, France\\
\email{crampe1977@gmail.com}
\and
Physics Department, PO Box 248046\\
University of Miami, Coral Gables, FL 33124 USA\\
\email{nepomechie@miami.edu}
\and
IVADO, Montr\'eal (Qu\'ebec), H2S 3H1, Canada
}
\maketitle  
% typeset the title of the contribution

\textit{Dedicated to Gordon Semenoff on the occasion of his 70th birthday}

\begin{abstract}
This paper offers a review of recent studies on the entanglement of free-fermion systems on graphs that take advantage of methods pertaining to signal processing and algebraic combinatorics. On the one hand, a parallel with time and band limiting problems is used to obtain a tridiagonal matrix commuting with the chopped correlation matrix in bispectral situations and on the other, the irreducible decomposition of the Terwilliger algebra arising in the context of $P$-polynomial association schemes is seen to yield a simplifying framework.

% We would like to encourage you to list your keywords within
% the abstract section using the \keywords{...} command.
\end{abstract}
\section{Introduction}
Quantifying the entanglement of quantum many-body problems is a meaningful issue and free-fermion systems provide fertile ground for such studies. Roughly speaking, %Grosso modo, 
the model is taken in its ground state, split in two parts and the entanglement of one part relative to the other is explored. This review %The paper 
offers a survey of some papers written on the subject by the authors and collaborators over the recent years \cite{cra:nep:vin, cram:nep:vin, ber:cra:vin, bern:cra:vin, cra:guo:vin, ber:cra:nep:par:pou:vin,par:ber:cra:vin, ber:man:par:vin}.

\section{Free fermions on weighted paths}
We first consider  
fermionic chains with dynamics described as follows.
\subsection{The Hamiltonian and its eigenstates}
The system is an open chain of length $N+1$ described by the Hamiltonian
\begin{equation}
\label{eq:H}
    H=\sum_{n=0}^{N-1}J_n(c_{n+1}^{\dagger}c_n + c_n^{\dagger}c_{n+1}) - \sum_{n=0}^N B_n c_n^{\dagger} c_n, \quad J_{-1} = 0,
\end{equation}
with the constants $J_n$ and $B_n$ real and positive and $c_n$,  $c_n^{\dagger}$ the fermionic operators obeying the anticommutation relations:
\begin{equation}
    \{ c_m, c_n^{\dagger} \} = \delta_{m,n},  \qquad \{c_m , c_n \} = 0, \qquad m,n=0,\dots N.
\end{equation}
Let $\{ |n\rangle, \ n=0,\dots,N\}$ be the orthonormal \textit{position basis} made out of the characteristic vectors $|n\rangle$ of $\mathbb{C}^{N+1}$. Introducing the matrix $\Lambda$ defined by
\begin{equation}
    \Lambda |n\rangle = J_{n-1}|n-1\rangle - B_n|n\rangle+J_n|n+1\rangle  ,
    \label{eq:a}
\end{equation}
the Hamiltonian $H$ can be written in the form: $ H=\sum_{m,n=0}^N \Lambda_{mn}c_m^{\dagger}c_n$ with $\Lambda_{mn}=\langle m|\Lambda |n\rangle $. Note that $\Lambda$ can be viewed as the adjacency matrix of a weighted path with self-loops.

The eigenstates of $H$ are obtained by diagonalizing $\Lambda$. This brings the orthonormal \textit{energy basis} of $\mathbb{C}^{N+1}$:
\begin{equation}
    \{|\omega _k\rangle,\ k=0,\dots,N \quad|\quad \Lambda |\omega _k \rangle = \omega _k |\omega _k\rangle \}.
   \label{eq:b}
\end{equation}
We order the energies as $\omega_k \leqslant \omega_{k+1}$. Owing to \eqref{eq:a} and \eqref{eq:b}, the wavefunctions $\phi _n(k) = \langle n| \omega_k\rangle$ are expressed in terms of orthogonal polynomials of a discrete variable. Under the canonical transformation $\tilde c_k = \sum_{n=0}^N \phi_n(k) c_n$, the Hamiltonian is brought in the form $H = \sum_{k=0}^N \omega _k \tilde c _k^{\dagger}\tilde c _k$. With $|0\rangle \! \rangle$ the vacuum state defined by the property $c_n |0\rangle \! \rangle = 0$ for $n= 0,\dots, N$, we readily have
\begin{equation}
    H\tilde c_{k_1}^{\dagger}\dots\tilde c_{k_m}^{\dagger} |0\rangle \! \rangle = \Big(\sum_{i=1}^m \omega_{k_i}\Big) \tilde c_{k_1}^{\dagger}\dots\tilde c_{k_m}^{\dagger} |0\rangle \! \rangle.\
\end{equation}

\subsection{Correlations and entanglement}

The ground state $|\Psi_0\rangle \! \rangle$ is obtained by filling the Fermi sea, that is by populating the vacuum with the excitations of energies up to $\omega_K <   \omega_N$ : $|\Psi_0\rangle \! \rangle = \tilde c_0^{\dagger}\dots\tilde c_K^{\dagger} |0\rangle \! \rangle $.

The correlation matrix has elements
\begin{equation}
   \bar C_{mn} = \langle \! \langle \Psi _0| c_m^{\dagger} c_n |\Psi_0\rangle \! \rangle = \sum_{k=0}^K \phi_m(\omega _k) \phi_n(\omega _k), \quad m,n = 0,\dots, N, 
\end{equation}
thus showing that $\bar C = \sum_{k=0}^K |\omega _k\rangle \langle \omega _k|$, in other words that the correlation matrix $ \bar C$ is the projector $\Pi _E$ on the subspace of energy states in the Fermi sea.

In the following, we discuss entanglement between two complementary parts of the chain. 
Part 1 
consists of the sites $ \{0, 1, \dots, \ell \}$, and part 2, the complement of part 1, is formed of the sites $ \{\ell +1, \dots, N\}$. With the full system in the ground state and thus described by the density matrix $\rho = |\Psi _0 \rangle \! \rangle \langle \! \langle \Psi _0 |$, the entanglement of part 1 with part 2 is completely accounted for by the reduced density matrix $\rho_1 = \Tr_2 |\Psi _0 \rangle \! \rangle \langle \! \langle \Psi _0 | $, where the trace is over the subspace of the Fock space generated by the creation operators $c^{\dagger} _i$ associated to part 2, i.e., with $i \in \{\ell +1, \dots, N\}$.

In the case of free-fermion chains, a considerable simplification \cite{pes} known as the Peschel trick occurs; namely the reduced density matrix, here a $2^{\ell + 1} \times 2^{\ell + 1}$ matrix, can be obtained  from the chopped correlation matrix $C$. This matrix is the restriction of the full correlation matrix $\bar C$ to the subspace $S \subset \mathbb C^{N+1}$ spanned by the vectors $\{ |0\rangle, \dots, |\ell \rangle \}$ and thus, a $(\ell + 1) \times (\ell + 1)$ matrix. If we denote by $\Pi _S $ the space projector on $S$, in view of the observation made before about $\bar C$, we arrive at the conclusion that the chopped correlation matrix $C$ is the key entity and that it is given by the product of three projectors: $C = \Pi _S \Pi _E \Pi _S$. Indeed, the (von Neuman) entanglement entropy $\mathfrak S = - {\rm \Tr}\, \rho _1 \log\, \rho _1$ is %seen to be 
given by \cite{pes}
\begin{equation}
    \mathfrak S = - {\rm Tr} \big[ C\, \log \, C 
    + (1 - C)\, \log\, (1 - C)\big].
\end{equation}

Intuitively, the entanglement entropy $\mathfrak S$ measures how mixed the reduced density matrix $\rho _1$ is. In a system with no entanglement, $\rho _1$ would simply be a projector and the entropy is trivially zero.

\subsection{A commuting (Heun) operator}

To compute the entanglement entropy $\mathfrak S$ we hence need to diagonalize the chopped correlation matrix $C = \Pi _S \Pi _E \Pi _S$. As a rule, this is a full matrix with eigenvalues near 0, a feature that does not facilitate a numerical treatment. We here wish to stress a useful observation, namely that for a class of fermionic chains, there exists a tridiagonal matrix $T$ with a well-behaved spectrum and such that $[T, C] = 0$, see \cite{eis:pes, cra:nep:vin, cram:nep:vin}.

Such occurrences are particularly opportune because $T$ shares its eigenvectors with $C$ and is easier to diagonalize (numerically). We shall now discuss situations when such a commuting operator is present and indicate how it can be obtained. 

The key is bispectrality. If the constants $J_n$ and $B_n$ in the Hamiltonian $H$ are such that the orthogonal polynomials arising in the wavefunctions $\phi_n(k)$ belong to the Askey scheme \cite{koe:les:swa}, these wavefunctions are bispectral. (Note that there are many such choices.) In these cases, there is an operator $X$ on $\mathbb C^{N+1}$ that is diagonal in the position basis and tridiagonal in the energy basis; that is, there is an $X$ such that
\begin{equation}
    X|n\rangle = \lambda _n |n\rangle, \qquad X|\omega _k \rangle = \bar J _{k-1} |\omega _{k-1}\rangle - \bar B _k |\omega _k\rangle + \bar J _k |\omega _{k+1}\rangle.
\end{equation}
This follows from the fact that $\phi _n(k) = \langle n|\omega _k\rangle$, being bispectral, obeys a difference equation of the form
\begin{equation}
    \lambda _n \phi _n(k) = \bar J _{n-1} \phi _n (k-1) - \bar B _k \phi _n(k) + \bar J _k \phi _n(k+1),
\end{equation}
in addition to the three term recurrence relation that is implied by the reciprocal action of the operator $\Lambda$ in the two bases:
\begin{equation}
  \Lambda |\omega _k\rangle = \omega _k |\omega _k \rangle, \qquad \Lambda |n\rangle = J_{n-1} |n - 1\rangle - B_n |n\rangle + J_n |n+1\rangle.  
\end{equation}

Under such circumstances, the commuting operator can be obtained as follows. To all bispectral problems, one may associate a so-called algebraic Heun operator defined as the most general bilinear expression in the two bispectral operators \cite{gru:vin:zhe}. For simplicity, we shall here consider a special case and the operator 
\begin{equation}
    \bar T = \{X, \Lambda\} + \mu X + \nu \Lambda,
\end{equation}
where the scalars $\mu$ and $\nu$ are for the moment unspecified. Clearly, $\bar T$ is tridiagonal in both the position and the energy bases. From the specific action of $\bar T$ in these bases, it is not difficult to see that by choosing for $\mu$ and $\nu$ the values
\begin{equation}
    \mu = -(\omega_K + \omega_{K+1}), \quad %{\rm and}
    \quad \nu = -(\lambda_{\ell} + \lambda_{\ell + 1}),
\end{equation}
$\bar T$ preserves the subspace S and the one spanned by the energy eigenvectors belonging to the Fermi sea. With T the restriction to $S$ of $\bar T$, it follows that $[T, \Pi _S] = [T, \Pi _E] = 0$ and hence that $[T, C] = 0$ since $C = \Pi _S \Pi _E \Pi _S$.

This result stems from a striking parallel \cite{gio:kli, eisl:pes} with the celebrated treatment \cite{sle} by Slepian et al. of the time and band limiting problem in signal processing. This deserves a short digression. The central and prototypical question is: how to best concentrate in a time interval $-T\le t \le T$ a signal $f(t)$ which is limited to a frequency band $[-W, W]$? This is in principle answered by looking for the eigenfunctions of the integral operator $G$ with the sinc kernel, namely by solving
\begin{equation}
    GF (p) = \int _{-W}^W dp^\prime \frac{\sin (p-p^\prime) T}{\pi (p - p^\prime)}F(p^\prime) = \lambda F(p). 
\end{equation}
Note that $G$ can also be written as the product of three projector: $G=\Pi _W^p \widetilde \Pi _T^p \Pi _W^p$ with $\Pi _L^x g(x) = [\Theta (x+L) - \Theta (x- L)] g(x)$, $\widetilde \Pi _T^p$ the Fourier transform of $\Pi _T^t$ and $\Theta(x)$ the Heavyside function.

This should be the end of the story but G, a non-local operator, proves also intractable numerically. The way out came from the remarkable discovery \cite{sle} that the spheroidal wave operator 
\begin{equation}
    D = \frac{1}{2} \Big\{\frac{d^2}{dp^2}, p^2\Big\} - W^2\frac{d^2}{dp^2} + T^2p^2\
\end{equation}
commutes with G. This result can be obtained \cite{gru:vin:zhe} from the rather obvious observation that the Fourier function $e^{ipx}$ are solutions to a simple bispectral problem with $\frac{d^2}{d p^2}$ and $p^2$ playing the roles of $\Lambda$ and $X$ in the scenario described before. The parameters of the Heun operator are here also fixed by demanding that the commutators with the projectors $\Pi _W^p$ and $\widetilde \Pi _T^p$ be equal to 0.

The parallel between the entanglement analysis of free-fermion chains and the band and time limiting problem is thus quite clear. The filling of the Fermi sea in the construction of the ground state corresponds to the band limiting. Splitting space in two parts and restricting to one is akin to time limiting. The main problem regarding the fermionic chain is to diagonalize the chopped correlation matrix $C = \Pi _S \Pi _E \Pi _S$ while in the band and time limiting case, it is to diagonalize the integral operator $G=\Pi _W^p \widetilde \Pi _T^p \Pi _W^p$. The understanding that the existence of a commuting operator is rooted in underlying bispectral problems leads then in both contexts, to a simple identification of this operator from the respective algebraic Heun operators.

\subsection{A non-homogeneous example: the Krawtchouk chain}

A nice example of a chain with bispectral features arises for the Hamiltonian \eqref{eq:H} with the following choice of parameters:
\begin{equation}
    J_n = \sqrt{(N-n)(n+1)p(1-p)}, \quad B_n = - (Np - n(1-2p)), \quad p \in [0,1].
    \label{eq:c}
\end{equation}
In this case  $\omega_k = \lambda_k = k$ and the wavefunctions are given as follows in terms of Krawtchouk polynomials (the ${_2}F_1$ part in the formula below) \cite{koe:les:swa}:
\begin{equation}
  \phi_n(k) = (-1)^n \sqrt{p^{n+k} (1-p)^{N-n-k} \binom{N}{n} \binom{N}{k} }
\; {_2}F_1 \; \left ({-n, -k \atop -N} ; \frac{1}{p} \right).
\end{equation}

Note that $B_n$ is a constant for $p=1/2$. In this case, 
the non-zero matrix elements $T_{mn}=T_{nm}$ of the symmetric commuting operator are given by \cite{cra:nep:vin}:

\begin{equation}
     T_{n,n} = \frac{N}{2}(2n-2\ell-1)-n(2K+1), \quad T_{n-1,n} = (n-\ell -1) \sqrt{n(N-n+1)}.
\end{equation}

As for the entanglement entropy for the half chain and at half filling it is well approximated by\footnote{Note that in \eqref{eq:SvnKr} we fixed a small sign typo from the original publication \cite{ber:cra:nep:par:pou:vin}.} \cite{ber:cra:nep:par:pou:vin}:
\begin{equation}
\label{eq:SvnKr}
    \mathfrak S = \frac{1}{6} \log\frac{N+1}{2} + a(p) - \frac{1}{2(N+1)}\frac{\cos(\frac{\pi}{2}\frac{N+1}{m(p)})}{\sin(\frac{\pi}{2m(p)})} + \dots
\end{equation}
where 
\begin{equation}
    m(p)=\frac{1}{2}\left(1-\log p +\frac{1-\log 2}{2p}\right) \,,
\end{equation}
and $a(p)$ is a non-universal constant with respect to $N$. 

\section{Free fermions on graphs}

We now consider free fermions on higher dimensional non-oriented graphs. Let $V= \{v_0,\dots,v_D\}$ be the set of vertices and $E \subset V \times V$, the set of edges. The orthonormal canonical basis $\{|v_0\rangle,\dots, |v_D\rangle \}$ will be referred to as the position basis with the vector $|v_i\rangle$ associated to the vertex $v_i$. The $(D+1) \times (D + 1)$ symmetric adjacency matrix $A$ has entries given by 
\begin{equation}
    A_{ij} = \langle v_i|A|v_j \rangle =
    \begin{cases}
        1 \quad \text{if} \quad (v_i, v_j) \in E\\
        0 \quad \text{otherwise}.
    \end{cases}
\end{equation}
The Hamiltonian $\mathcal H$ is taken to be $\mathcal H= \sum_{m,n = 0}^D A_{m,n} c_m^{\dagger} c_n$ where $c_n$ and $c_n^{\dagger}$ are the usual fermionic operators at $v_n$. Thus the adjacency matrix (made out of only $0$s and ${1}$s) plays a role similar to the tridiagonal matrix $\Lambda$ introduced in the description of fermionic chains. 
We shall focus below on a natural family of graphs.

\subsection{The hypercube and the Krawtchouk chain}\label{sec:KrawChain}

In the case of the hypercube $Q_N$ in $N$ dimensions, $V = \{0,1\}^{\otimes N}$; that is, the vertices are strings of $N$ bits. Two vertices $v_i, v_j \in V$ are linked if they differ by only one entry, i.e., if they are at Hamming distance $d(v_i, v_j) = 1$. The case $N=1$ corresponds to the complete graph (where all vertices are connected to one another) with two vertices $K_2$. In general, $Q_N$ is the $N$-fold Cartesian product of $K_2$, i.e. $Q_N = (K_2)^{\square N}$.

Let us now establish a connection between the system consisting of fermions on the hypercube and the Krawtchouk chain. Pick $0 = (0, \dots, 0)$ as a reference point on $Q_N$. Organize $V$ in columns $V_n = \{x \in V \;|\; d(0,x)=n\}$ made out of all vertices at distance $n = 0, \dots , N $ of $0$.  It is easy to see that $k_n = \text{Card} (V_n) = \binom{N}{n}$. Let us label the vertices in the column $V_n$ by $V_{nm}, m= 1, \dots, k_n$ and form the $n$-qbit Dicke or column vector states:
\begin{equation}
    |col \; n\rangle = \frac{1}{\sqrt{k_n}} \sum_{m=1}^{k_n} |V_{nm}\rangle.
\end{equation}
It is easy to see \cite{chr:dat:eke:lan, ber:cha:lor:tam:vin} that $\langle col \; n+1|\;A\:|col \; n\rangle = \sqrt{(n+1)(N-n)}$ which is equal to  twice the expression of $J_n$ for the Krawtchouk chain when $p=\frac{1}{2}$ as per~\eqref{eq:c}. In other words, for $\Lambda$ corresponding to the Krawtchouk chain with $p=\frac{1}{2}$, we have $\langle col \; n+1|\:A\;|col \: n\rangle = 2\; \langle n+1|\;\Lambda \: |n \rangle $. When $p=\frac{1}{2}$, the diagonal term in the Hamiltonian of the Krawtchouk chain does not depend on $n$ and thus yields a global constant that can be subtracted; we hence find that up to an overall multiplicative factor the hypercube system projects to the $p=\frac{1}{2}$ Krawtchouk chain. The reason for this is that $Q_N$ is a distance-regular graph, implying that each vertex in column $V_n$ is connected to the same number of vertices in the column $V_{n+1}$ and vice versa. This observation suggests that the entanglement properties of free fermions on the hypercube bear a relation with those of the Krawtchouk chain. This connection 
can further be understood in terms of association schemes. The hypercube $Q_N$ is part of the family of graphs on $V$ consisting of those where it is the vertices at Hamming distance $0$, $1$, or $2$, up to $N$ that are connected by an edge. These are the graphs of the (binary) Hamming association scheme that have adjacency matrices whose action is closed on the space spanned by the column vector states.
We discuss these constructs more generally next.

\subsection{Association schemes}

An important concept in algebraic combinatorics is that of (symmetric) $d$-class association schemes \cite{ban:ban:ito:tan} which can be considered as ensembles of $d+1$ undirected graphs on a set of vertices $V$ with cardinality $|V|$ satisfying certain axioms. Such an ensemble of graphs may be seen as colorings of the edges of the complete graph with $d$ colors.  In terms of the corresponding adjacency matrices $A_i,\ i=0, \dots, d$, the axioms are equivalent to
\begin{equation}
  \sum_{i=0}^d A_i = J, \quad A_0 = I, \quad A_i = A_i^T, \quad A_iA_j = \sum_{k=0}^d p_{ij}^kA_k,  
\end{equation}
where $J$ is the all $1$ matrix, $I$ the identity and $p_{ij}^k$ integers referred to as the intersection numbers. The commutative $d+1$ algebra thus generated by the adjacency matrices is called the Bose-Mesner algebra. Since the symmetric adjacency matrices all commute, they can be diagonalized simultaneously and admit the spectral decomposition
\begin{equation}
    A_i=\sum_{j=0}^d \theta_i(j) E_j, \qquad E_i = \frac{1}{|V|} \sum_{j=0}^d \theta_i^*(j) A_j,
\end{equation}
with $E_j$ the idempotents projecting on the eigenspaces ($E_iE_j=\delta _{ij} E_i$ and $\sum_{i=0}^d E_i = I).$ Finally, it is known that the distance matrices of distance regular graphs lead in a one-to-one way to association schemes that are $P$-polynomial in that 
 $A_i = p_i(A_1)$ for $p_i$ a polynomial of degree $i$. We shall often assume this situation to hold in the following. Dually, an association scheme is called $Q$-polynomial if there is an ordering such that its primitive idempotents $E_i$ are given as polynomials of degree $i$ of $E_1$ (under the entry-wise product).

\subsection{The Terwilliger algebra and the correlation matrices}

An algebra introduced by Terwilliger \cite{ter} and extending the Bose-Mesner one can be attached to an association scheme and is relevant to our entanglement studies.  Its definition requires picking a reference vertex $v_0$ and introducing 
the dual matrices $A_i^*$ and $E_i^*$ as the diagonal matrices with entries:
\begin{equation}
    [A_i^*(v_0)]_{vv} = |V| [E_i]_{v_0v}, \qquad [E_i^*(v_0)] = [A_i]_{v_0v}.
\end{equation}
Note that $E_i^* E_j^* = \delta_{ij} E_i^*$ and that $E_i^*$ projects on the position subspace spanned by the vectors $|v\rangle$ corresponding to vertices connected to $v_0$ in the graph with adjacency matrix $A_i$, i.e., the column space at distance $i$ from $v_0$. Recall that $E_i$ projects on the energy eigenspaces of the adjacency matrices that intervene in the Hamiltonians. The Terwilliger algebra $\mathfrak T$ is generated by the adjacency matrices $\{A_0, \dots, A_d\}$ and their duals $\{A_0^*, \dots, A_d^*\}$ or equivalently by the sets of projectors $\{E_0, \dots,E_d\}$ and $\{E_0^*, \dots. E_d^*\}$.

It is appropriate to remark at this point that the entanglement analysis of fermions on graphs proceeds much as for fermion chains. The ground state is defined by filling the vacuum state $|0\rangle \! \rangle $ with the excitations corresponding to a subset $SE$ of the eigenvalues   $\theta (j)$ of the adjacency matrix A (or of combination) chosen for Hamiltonian. The correlation matrix $\bar C$ is given by 
\begin{equation}
    \bar C = \sum_j E_j = \Pi_{SE} \quad \text{for} \quad \theta(j) \in SE.
\end{equation}
The bipartition of the vertices (or positions) $V$ into $SV$ (part 1) and its complement (part 2) is typically be done by picking columns at the successive distances from $0$ to $\ell < d$ with respect to a vertex $v_0$. Since $E_i^* = \sum_{v | d(v_0,v)=i} |v\rangle\langle v|$, the projector on the position vectors of part 1 is $\Pi _{SV} = \sum_{i=0}^{\ell} E_i^*$ and the chopped correlation matrix thus reads $C= \Pi _{SV} \Pi _{SE} \Pi_{SV}$. It follows that this matrix $C$ that needs to be diagonalized actually represents an element of the Terwilliger algebra $\mathfrak T$, a point worth underscoring.

A natural strategy to carry out the entanglement analysis of fermions on graphs of ($P$- and $Q$- polynomial) association schemes thus presents itself: (i) Identify the Terwilliger algebra $\mathfrak T$ for the scheme; (ii) Decompose the regular representation of $\mathfrak T$ on $\mathbb C ^{|V|}$ into its irreducible components and as a result; (iii) Simplify the diagonalization of $C$ by working on irreducible subspaces. This last step can further be aided by the presence of a commuting operator belonging also to $\mathfrak T$. For $P$- and $Q$- polynomial schemes, $\mathfrak T$ is generated by $A_1 = A$ and $A_1^* = A^*$ solely as all the others matrices are polynomials of one or the other. The energy and position bases are respectively the eigenbases of $A$ and $A^*$. The action of each of these elements in the eigenbasis of the other is block tridiagonal. This implies that the overlaps between the two bases (the wavefunctions) are solutions of bispectral problems. Considering therefore the (generalized) algebraic Heun operator 
$\bar T= \{A,A^*\} + \mu A^* + \nu A $, it is possible to find $\mu$ and $\nu$ so that $[\bar T, \Pi_{SV}] = [\bar T, \Pi_{SE}] =0$ and hence $[\bar T, C]=0$.

\subsection{Entanglement on graphs of the (binary) Hamming scheme}

 Let us bring as an example the celebrated Hamming scheme $\{A_i\}_{0\leqslant i\leqslant d}$ which was referred to at the end of the subsection on the hypercube. The matrices $A_i$ are the distance matrices of the hypercube and map $|0\rangle$ to column vectors, i.e., $A_i |0\rangle = \sqrt{k_i} |col \; i\rangle$. This scheme is known to be $P$- and $Q$- polynomial with the self-dual Krawtchouk polynomials arising in the expression of the (dual) adjacency matrices in terms of $A = A_1$ and $A^* = A_1^*$. For a bipartition where part 1 = $\{x\in V$ s.t. $d(0,x) \leqslant \ell \}$, the projectors arising in $C$ are polynomials in these matrices,
\begin{equation}
    \Pi_{SE} = \sum_{k = 0}^{\lfloor d/2\rfloor} \prod_{\substack{j = 0\\
    j \neq k}}^d \frac{(2j - d - A)}{2(j-k)}, \quad     \Pi_{SV} = \sum_{k = 0}^{\ell} \prod_{\substack{j = 0\\
    j \neq k}}^d \frac{(2j - d - A^*)}{2(j-k)}.
\end{equation}
The two matrices $A$ and $A^*$ generate $\mathfrak T$ which we easily recognize to be the Lie algebra $\mathfrak{su}(2)$. By applying the definitions of the adjacency matrix and its dual for $K_2$, it is seen that:
\begin{equation}
    A= \begin{pmatrix}
        0&1\\
        1&0
    \end{pmatrix} = \sigma_x,  \qquad A^* =\begin{pmatrix}
        1&0\\
        0&-1
    \end{pmatrix} = \sigma_z,
\end{equation}
where $\sigma _x$ and $\sigma _z$ are the usual notation for the Pauli matrices.
Since as already noted, the hypercube $Q_d = (K_2)^{\square d}$ it follows that in this case:
\begin{equation}
    A=\sum_{i=0}^{d-1} \underbrace{I\otimes\dots\otimes I}_i \otimes\:\sigma _x \otimes I \dots \otimes I, \quad A^*=\sum_{i=0}^{d-1} \underbrace{I\otimes\dots\otimes I}_i \otimes\:\sigma _z \otimes I \dots \otimes I.
\end{equation}
This corresponds to the diagonal embedding of $\mathfrak{su}(2)$ into $\mathfrak{su}(2)^{\otimes d}$  with $A$ and $A^*$ resulting from $d-1$ applications of the coproduct. The underlying representation is of course very well known. As anticipated the irreducible decomposition reduces the entanglement characterization problem on the graph to a combination of Krawtchouk chains, one of which being the chain discussed in Section \ref{sec:KrawChain}.
 
\section{Conclusion}

The connections that have been highlighted between the entanglement analysis of fermionic systems, signal processing and algebraic combinatorics have been put to use in a variety of contexts but many avenues still remain to be explored. The entanglement of free fermions on the Hamming, Johnson, Hadamard and folded cube graphs have been examined \cite{ber:cra:vin, bern:cra:vin, cra:guo:vin, ber:man:par:vin}. It would certainly be worth determining what other schemes \cite{bro:coh:neu} such as the dual polar \cite{berna:cra:vin} or Grassmann ones entail. In some cases the irreducible decompositions of the regular representation of the Terwilliger algebras have not yet been spelled out. $P$-multivariate association schemes and their graph descriptions are currently generating much interest \cite{bern:cra:pou:vin:zai, ban:kur:zha:zhu, cra:vin:zai:zha, ber:cra:vin:zai:zha}. Examining beyond the initial studies \cite{ber:cra:nep:par:pou:vin} the entanglement of fermionic systems built upon those structures should warrant attention. Finally, the bearing of graph symmetries on the entanglement properties of free fermions on these graphs is certainly a question that we plan to study in the future.

\section*{Acknowledgments}
LV is grateful to the organizers of the conference at the CRM in honour of Gordon Semenoff for having invited him to be part of this celebration. 

PAB holds an Alexander-Graham-Bell scholarship from the Natural Sciences and Engineering Research Council of Canada (NSERC). NC is partially supported by the IRP AAPT of CNRS. RN enjoys funding from the National Science Foundation under Grant No. PHY 2310594. GP holds a FRQNT and a CRM–ISM postdoctoral fellowship, and acknowledges support from the Mathematical Physics Laboratory of the CRM. The research of LV is funded in part by a Discovery Grant from NSERC.


\begin{thebibliography}{6}
%

\bibitem{cra:nep:vin}
Cramp\'e, N. Nepomechie, R., Vinet, L.: Free-Fermion entanglement and orthogonal polynomials.
Journal of Statistical Mechanics 093101 (2019)

\bibitem{cram:nep:vin}
Cramp\'e, N., Nepomechie, R., Vinet, L.: Entanglement in Fermionic Chains and Bispectrality.
Reviews in Mathematical Physics 33(7) (2021)

\bibitem{ber:cra:vin}
Bernard, P.-A., Cramp\'e, N., Vinet, L.: Entanglement of free fermions on Hamming graphs.
Nuclear Physics B986 116061 (2023)

\bibitem{bern:cra:vin}
Bernard, P.-A., Cramp\'e, N., Vinet, L.: Entanglement of free fermions on Johnson graphs.
Journal of Mathematical Physics 64 061903 (2023)

\bibitem{cra:guo:vin}
Cramp\'e, N., Guo, K., Vinet, L.: Entanglement of Free Fermions on Hadamard Graphs.
Nuclear Physics B 360 115176 (2020)

\bibitem{ber:cra:nep:par:pou:vin}
Bernard, P.-A., Cramp\'e, N., Nepomechie, R., Parez, G., Poulain d'Andecy, L., Vinet, L.: Entanglement of inhomogeneous free fermions on hyperplane lattices.
Nuclear Physics B 984 115975 (2022)

\bibitem{par:ber:cra:vin}
Parez, G., Bernard, P.-A., Cramp\'e, N., Vinet, L.: Multipartite information of free fermions on Hamming graphs.
Nuclear Physics B 990 116157 (2023)

\bibitem{ber:man:par:vin}
Bernard, P.-A., Mann, Z., Parez, G., Vinet, L.: Absence of logarithmic enhancement in the entanglement scaling of free fermions on folded cubes.
Journal of Physics A: Math. Theor. 57 015002 (2024)

\bibitem{pes}
Peschel, I.: Calculation of reduced density matrices from correlation functions.
Journal of Physics A: Math. Gen. 36 L205 (2003)

\bibitem{eis:pes}
Eisler, V., Peschel, I.: Properties of the entanglement Hamiltonian for free-fermion chains.
Journal of Statistical Mechanics 104001 (2018)

\bibitem{koe:les:swa}
Koekoek, R. Lesky, P.A., Swarttouw, R.F.: Hypergeometric Orthogonal Polynomials and Their q-Analogues.
Springer (2010)

\bibitem{gru:vin:zhe}
Gr\"{u}nbaum, F. A., Vinet, L., Zhedanov, A.: Algebraic Heun Operator and Band-Time Limiting.
Communications in Mathematical Physics 364 1041 (2018)


\bibitem{gio:kli}
Gioev, D., Klich, I.: Entanglement Entropy of Fermions in Any Dimension and the Widom Conjecture.
Physical Review Letters 96 100503 (2006)

\bibitem{eisl:pes}
Eisler, V., Peschel, I.: Free-fermion entanglement and spheroidal functions.
Journal of Statistical Mechanics 04028 (2013)

\bibitem{sle}
Slepian, D.: Some comments on Fourier analysis, uncertainty and modeling.
SIAM Review 25 379 (1983)

\bibitem{chr:dat:eke:lan}
Christandl, M., Datta, N., Ekert, A., Landahl, A. J.: Perfect state transfer in quantum spin networks.
Physical Review Letters 92 187902 (2004)

\bibitem{ber:cha:lor:tam:vin}
Bernard, P.-A., Chan, A., Loranger, E., Tamon, C., Vinet, L.: A graph with fractional revival.
Physics Letters A 382 259 (2018)

\bibitem{ban:ban:ito:tan}
Bannai, E., Bannai, E., Ito, T., Tanaka, R.: Algebraic Combinatorics.
De Gruyter (2021)

\bibitem{ter}
Terwilliger, P.: The subconstituent algebra of an association scheme.
Journal of Algebraic Combinatorics (Part I) 1 363 (1992), (Part II) 2 73 (1993), (Part III) 2 177 (1993)

\bibitem{bro:coh:neu}
Brouwer, A. E., Cohen, A.M., Neumaier, A.: Distance-Regular Graphs.
Springer (1989)


\bibitem{berna:cra:vin}
Bernard, P.A., Cramp\'e, N., Vinet, L.: The Terwilliger of symplectic dual polar graphs, the subspace lattices and $\mathcal{U}_q(sl_2)$.
Discrete Mathematics 12 113169 (2023)

\bibitem{bern:cra:pou:vin:zai}
Bernard, P.A., Cramp\'e, N., Poulain d'Andecy, L. Vinet, L., Zaimi, M.: Bivariate $P$-polynomial association schemes.
arXiv:2212.10824 (2022)

\bibitem{ban:kur:zha:zhu}
Bannai, E., Kurihara, H., Zhao, D., Zhu, Y.: Multivariate $P$ and/or $Q$-polynomial association schemes.
arXiv:2305.00707v2 (2023)

\bibitem{cra:vin:zai:zha}
Cramp\'e, N., Vinet, L. Zaimi, M., Zhang, X.: A bivariate $Q$-polynomial structure for the non-binary Johnson scheme.
arXiv:2306.01882 (2023)

\bibitem{ber:cra:vin:zai:zha}
Bernard, P.-A., Cramp\'e, N., Vinet, L., Zaimi, M., Zhang, X.: $m$-distance-regular graphs and their relation to multivariate $P$-polynomial association schemes.
arXiv:2309.16016

\end{thebibliography}
\end{document}